# Thermoresponsive PNIPAM/FeRh smart composite activated by magnetic field for doxorubicin release.

Abdulkarim A. Amirov[+*], Elizaveta S. Permyakova[+], Dibir M. Yusupov[∥], Irina V. Savintseva[±], Eldar K. Murliev[∥], Kamil Sh. Rabadanov[∥], Anton L. Popov[±], Alisa M. Chirkova[‡], Akhmed M. Aliev[∥]

[+]National University of Science and Technology MISiS, 119049, Moscow, Russia

[∥]Amirkhanov Institute of Physics of Dagestan Federal Research Center, Russian Academy of Sciences, 367003, Makhachkala, Russia

[±]Institute of Theoretical and Experimental Biophysics of the Russian Academy of Sciences, 142290, Pushchino, Russia

[‡]Hochschule Bielefeld University of Applied Sciences and Arts, Bielefeld, Germany

**ABSTRACT:** The ability to control by physical properties of the thermoresponsive polymer of PNIPAM by magnetocaloric effect was demonstrated by in-situ experiments on PNIPAM/FeRh smart composite. The concept of drug release loaded in smart composite by applying of 3 T magnetic field was demonstrated using an example of doxorubicin. Released results of the magnetic field applying the drug were detected using IV-VIS and Raman spectroscopy. In vitro, studies have demonstrated a high degree of PNIPAM/FeRh scaffold biocompatibility for primary mouse embryonic fibroblasts (PMEF) cell culture. PMEFs effectively adhered to the PNIPAM/FeRh scaffold surface and showed high metabolic and proliferative activity for 72 h after seeding.

## INTRODUCTION

Current trends in materials science for biomedical applications (drug delivery, cell engineering, regenerative medicine) are aimed at the design and development of new functional materials that significantly change their physical properties under the influence of external stimuli of different nature (magnetic field, electric field, pressure, temperature, UV light) [1]. Such a class of materials is known under the unifying term "smart", among which thermosensitive polymers [2–4] are one of the most promising ones for the development of controlled theranostics systems. The most common among them are water-soluble polymers that exhibit a lower critical solution temperature (LCST) or an upper critical solution temperature (UCST) in solutions. Polymers with LCST dissolve at lower temperatures, and heating their solutions leads to their phase separation. Among the known thermosensitive polymers with LCST, poly(N-isopropylacrylamide) (PNIPAM) is the most promising and was synthesized for the first time in the 1950s [5]. PNIPAM exhibits LCST in aqueous solutions and, when heated in water above 32 °C, undergoes a reversible transition from a swollen hydrated state to a wrinkled dehydrated state, losing about 90% of its volume[6]. The LCST temperature of PNIPAM is close to physiological temperatures, and due to this, it is considered a promising material for tissue engineering ("smart" scaffolds), regenerative medicine ("smart" coatings for implants), and theranostics (controlled drug release) [7–9]. Comparatively recently, a new method of manufacturing "smart" coatings for implants has been proposed in which the properties of thermosensitive polymers are controlled through the magnetocaloric effect (MCE), and the Fe-Rh alloy with a "giant" MCE in the region of physiological human temperatures was proposed as a control material [10,11]. The proposed concept combines two phenomena:

1) MCE, the essence of which is to change the temperature of a magnetic material under adiabatic conditions when the external magnetic field changes;

2) Phase transition with LCST, when the polymer reversibly changes its properties under the action of temperature.

In this case, the reversible transfer of the thermoresponsive polymer properties deposited on the substrate of magnetic material is carried out through a change in its temperature as a result of MCE when the magnetic field is switched on or off. The novelty of this approach lies in the fact that LCST is achieved noninvasively by applying an external magnetic field and without the use of traditional heating (cooling) systems requiring the supply of electrical wires, which is relevant for biomedical applications. In a paper [12], the proposed idea was tested by showing that a magnetic field could be used to control the properties of a thermosensitive polymer in a way that could be reversed. This was done using the example of a composite made of a PNIPAM layer deposited on a substrate of a classic magnetocaloric Gd material and a specially designed experimental insert in the *in situ* mode. However, the results presented in [12] demonstrate only the workability of the idea, and the results are far from the possibilities of practical applications. In addition, Gd is a biologically toxic material, and the magnetic field used to activate the thermosensitive polymer is 8 T, which exceeds the parameters of magnetic field sources used in modern clinical medicine (e.g., magnetic tomographs). The temperature of the magnetic phase transition Gd, which is the "working point" of the composite, is in the region of room temperature (20 ºC and does not allow for effective control of the properties of the polymer PNIPAM with a LCST of about 32 ºC. The present study is devoted to the development of these approaches using less biologically toxic magnetocaloric materials with magnetic phase transition temperatures close to human physiological temperatures and to the search for effective solutions for drug release from such composites through a magnetic field. The present study is devoted to the development of these approaches with the use of less biologically toxic magnetocaloric materials with temperatures of magnetic phase transitions close to physiological human temperatures and the search for effective solutions of drug release from such composites through a magnetic field. Taking this into account, as a magnetocaloric material, alloys of the FeRh family with compositions close to equiatomic have been proposed, which exhibit record MCE in the region of human physiological temperatures [13,14] and, according to the reported data do not exhibit high cytotoxicity [15].

**EXPERIMENTAL SECTION**

**Materials.** PNIMAM ($M_n$~40 000, Alfa Aesar, Kandel, Germany), Doxorubicin (50 mg, Doxorubicin-LANS®, LENS-PHARM LLC, Russia), Fe (99.98%) and Rh (99.8%), Ethanol (99,7 %, VECTON, Russia).

**Sample Preparation.** The experimental part of the work was subdivided into two phases: the first stage was devoted to the verification of the concept of controlling the properties of a thermosensitive polymer through a magnetic field according to the protocol used in [12] and the second stage is closer to practical applications and is aimed at finding solutions to demonstrate the possibility of discharging real drugs using a magnetic field, the magnitude of which is achievable on commercial MRI tomographs.

Moving to the second stage of the experiment assumes successful proof of concept during the execution of the first stage. With this in mind, we fabricated two types of composite:

1) a classical one, which was a film of thermosensitive polymer PNIPAM deposited on a FeRh alloy substrate, and

2) a modified one, which was a patterned surface of a FeRh alloy substrate on which a drug was loaded and coated with a layer of thermosensitive polymer film PNIPAM.

The PNIPAM/FeRh composite sample was obtained by depositing PNIPAM polymer on the substrate of the $Fe_{49}Rh_{51}$ (FeRh) alloy sample using the solvent casting method by pouring the polymer solution on the substrate [16]. Arc melting high-purity Fe and Rh powders in a helium atmosphere at $10^{-4}$ mbar pressure, then annealing at 1000 °C for 7 days and air quenching, produced the $Fe_{49}Rh_{51}$ sample that served as the substrate. The $Fe_{49}Rh_{51}$ sample used as a substrate had a plate shape with dimensions of 4.6×3.8×0.2 mm$^3$ and was studied in detail in [17,18]. The process for creating the polymer backbone involved mixing PNIMAM powder in ethanol at room temperature until it dissolved. The 3% PNIPAM solution so made was then uniformly distributed and leveled on the FeRh substrate using the doctor's blade technique described in [19,20]. Then the sample was slowly dried overnight at 22 ºC in a closed dish to reduce porosity. The resulting sample was then re-dried at 60 ºC for 2 hours. It is worth noting that some of the key challenges in realizing magnetic field-controlled drug release approaches from a composite of magnetocaloric material and thermosensitive polymer are:

1. Losses associated with heat dissipation on the composite's surrounding medium, including biological, in the case of their use in real-life applications.
2. Adhesion of the polymer under cyclic exposure to a magnetic field, there is a high probability of polymer separation from the substrate. It is necessary to develop various physical and chemical methods of surface modification (plasma treatment, laser patterning, polymer "crosslinking", etc.).
3. Qualitative and quantitative assessment of the drugs released with the help of magnetic fields actually used in biomedicine.

From the above-mentioned points, the problems of heat dissipation research in real biological systems are, in our opinion, more studied and have been studied in a number of works [21,22]. For example, such a problem related to heat dissipation on biological objects is actively investigated in the problems of magnetic hyperthermia [23]. In the works [24], drug loading by its distribution in the volume of an aqueous solution of polymer PNIPAM in a hydrated state was proposed. As an alternative solution to optimize drug loading and discharge, we proposed modification of the FeRh surface by fabricating on its surface periodic structures like "wells" into which the drug is loaded, which in turn are coated with a layer of thermosensitive polymer. This approach, according to our assumption, will solve two problems: 1) loading a larger amount of drug; and 2) improving the adhesion of the polymer by changing the increase in the surface area of FeRh.

The composite sample PNIPAM/DOX/FeRh for the second part of the experiment was a FeRh alloy sample on whose modified surface the drug doxorubicin (DOX) was loaded and coated with a film of the thermosensitive polymer PNIPAM. For this purpose, using a fiber laser engraver (50 W power), "wells" were made on the entire surface of the FeRh sample at a fixed distance from each other). The resulting FeRh sample was washed in ethanol and placed in an ultrasonic bath for 30 minutes to further clean the surface. Drug loading and subsequent coating with the thermosensitive polymer PNIPAM were carried out using a similar protocol to that used for the first sample. Drug loading and subsequent coating with the PNIPAM polymer were carried out using a similar protocol to the first sample using the doctor's blade technique. Doxorubicin (DOX) was used as a drug. DOX belongs to the pharmacotherapeutic group of antitumor agents and is a cytotoxic anthracycline antibiotic isolated from Streptomyces peucetius var. Caesius culture is used for the treatment of a number of malignant neoplasms (solid tumors, leukemia, and lymphoma) [25]. The choice of this drug was primarily due to its availability and ease of detection using UV-visible spectroscopy techniques.

**Measurement Techniques.** The microstructure of the samples was looked at with a LEO-1450 scanning electron microscope (Leica Micro-systems Wetzlar Gmbh, Wetzlar, Germany) with an attachment for elemental composition analysis based on energy dispersive spectroscopy. The MCE was studied using the direct method described in [14,26]. To observe the property control of composite samples via magnetic field in the in situ mode, the experimental insert described in [12] was used. A cryogenic closed-loop magnetic system with a maximum field in the working region of 8 T was used as a magnetic field source. The insertion into the magnetic field and withdrawal from the magnetic field of the measuring insert were carried out using a linear actuator. A Shimadzu UV-3600 UV spectrophotometer was used to detect the doxorubicin release. Spectra registration conditions: slit 1.0 nm; single scan mode; medium scanning speed; quartz cuvette (1 cm); registration range 300–1200 nm. Data processing program UV Probe 2.10 Raman spectra were recorded in backscattering geometry on a Senterra confocal Raman microscope (Bruker, Germany). Raman spectra were measured at room temperature with laser excitation (532 nm), spectral measurement range 50–3500 $cm^{-1}$ with a resolution of 3-5 $cm^{-1}$, integration time of each scan 20 s, and number of scans 20. The laser is focused on the sample using a 50x objective. Low laser power was used to avoid overheating effects due to laser radiation.

**Cell culture.** We used primary mouse embryonic fibroblasts (PMEF) at the 3rd passage, harvested from the 13-day-old embryos of the B10GFP/Balb/c hybrid mice carrying a green fluorescent protein (GFP). Employing this type of cell culture was presumed expedient because one needed to avoid applying fluorescent dyes in order to prevent the coloration of the scaffolds under study. The mice were received from the vivarium of the Institute of Cell Biophysics RAS (Puchshino). The experiments were approved by the ethics committee of the Institute of Theoretical and Experimental Biophysics for the care and use of laboratory animals. The experiments conformed to the regulations and legal acts concerning the procedures of animal experiments and the humane treatment of animals (in accordance with Directive 2010/63/EU of the European Parliament and of the Council of September 22, 2010 on the Protection of Animals Used for Scientific Purposes). Mice were euthanized by cervical dislocation on the 13th day of gestation. Their uteri with embryos were isolated. After tissue dissociation in the 0.25% trypsin-0.02% ETDA solution (PanEco, Russia) for 30 min at 37 °C, the cells obtained were collected with centrifugation for 2 min at 1500 rpm. Then, the cells were resuspended to get a single-cell state in a milieu of DMEM/F12 (PanEco, Russia) with 10% fetal bovine serum (FBS) (in a 1:1 ratio). The resulting suspensions were transferred into vials (25 $cm^2$) and cultivated in a 5% $CO_2$ atmosphere at 37.0 C in DMEM containing 10% FBS (HyClone, USA), 100 units/ml of penicillin/streptomycin, and the addition of 2 mML glutamine. When the subconfluent state was reached, the cells were treated with a 0.25% trypsin-EDTA solution and seeded into vials (75 cm2) in a 1:3 ratio. The cultivation was carried out in a DMEM/F12 medium containing 10% fetal bovine serum (HyClone, USA), 100 units/ml of penicillin/streptomycin, and the addition of 2 mM glutamine. Cells from 3–4 passages were used in our experiments. After 24, 48, and 72 hours of cell cultivation on the polymer scaffold, the cell cultures were photographed with a BioRad Zoe Fluorescent Imager.

## RESULTS AND DISCUSSION

SEM images of the PNIPAM/FeRh composite sample are shown in Figure 1. The thickness of the PNIPAM polymer layer on the FeRh substrate, estimated from the SEM image taken in profile, was about 20 µm (Fig. 1 a). A better uniformity of the polymer layer and its adhesion were observed for the PNIPAM/FeRh sample compared to the PNIPAM/Gd composite investigated in [12].

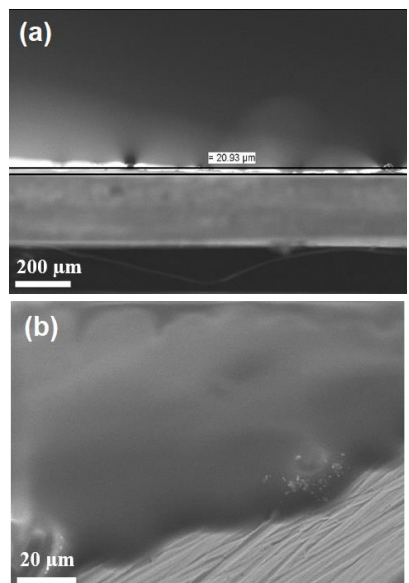

**Figure 1.** SEM images of the PNIPAM/FeRh composite sample taken in profile (a) and from the surface (b).

This is probably due to the presence of microscratches on the FeRh surface obtained as a result of its grinding (Figure 1b). The structure and surface of the PNIPAM/DOX/FeRh sample were investigated before and after coating with a drug-preloaded polymer. SEM images obtained for the laser-modified FeRh surface showed that the obtained "wells" have a circular shape with an average diameter of about 180 µm and are about 150 µm apart (Figure 2).

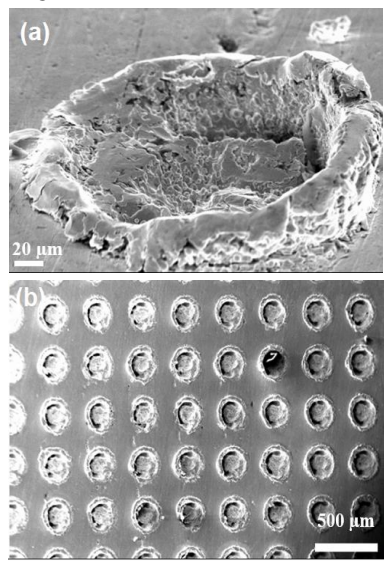

**Figure 2.** SEM images of the FeRh sample after laser treatment: separate "well" (a) and surface (b).

The shape of the "wells" in general is close to a hemisphere, with inhomogeneity of the inner surface and characteristic melting of the edges as a result of laser irradiation (Figure 2a). The exact determination of the depth of the "wells" based on the obtained SEM images is problematic; their maximum depth (taking into account the "rim" formed as a result of the impact of the laser beam along the entire diameter of the "well") is in the range of 40–80 µm. Figure 3 shows optical (b) and SEM (b-d) images of the PNIPAM/DOX/FeRh composite sample.

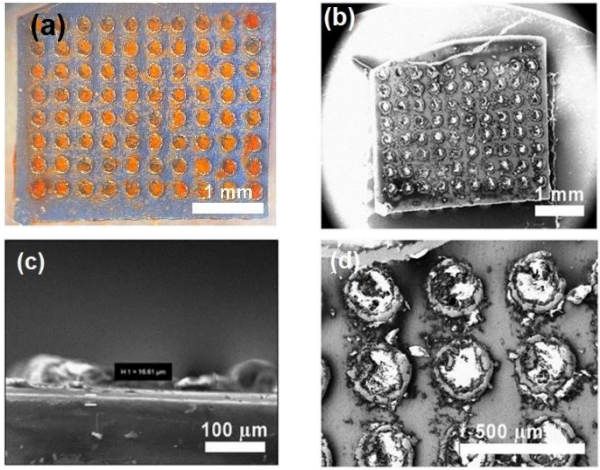

**Figure 3.** Optical (a) and SEM (b-d) images of the PNIPAM/DOX/FeRh sample.

SEM images of the surface and profile of the sample show that the polymer PNIPAM does not fully cover the surface of the FeRh substrate with a smooth layer, but it completely cover the "wells" loaded by doxorubicin. The thickness of the polymer layer at the ends of the FeRh plate is about 16 μm. The loading of doxorubicin into the "wells" is also visible by contrast on SEM images.

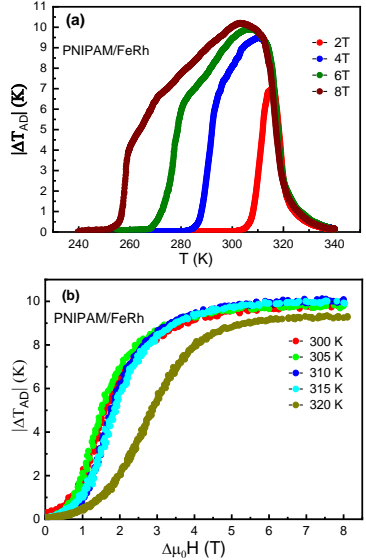

**Figure 4.** Temperature (a) and field dependences of the adiabatic temperature change of the PNIPAM/FeRh sample in magnetic fields up to 8 T.

Prior to the experiments to demonstrate the possibility of controlling the properties of a thermosensitive polymer by a magnetic field, preliminary MCE measurements were carried out on the PNIPAM/FeRh composite.

Fig. 4 shows the temperature dependences of the adiabatic temperature change of the samples $|\Delta T_{AD}|$ measured at different values of the applied magnetic field in the range of 0–8 T. The temperature dependences of $|\Delta T_{AD}|$ for PNIPAM/FeRh show a characteristic behavior for $Fe_{49}Rh_{51}$ alloy with a maximum of 7 K at 2 T in the region of 315 K, which broadens towards low temperatures with increasing field [14,17,27,28]. The values of $|\Delta T_{AD}|$ for PNIPAM/FeRh are lower than for pure $Fe_{49}Rh_{51}$ alloy, which is due to heat losses on the polymer layer of PNIPAM, which in this case is passive and does not participate in MCE

[13,14,17]. The field dependences of $|\Delta T_{AD}|$ obtained at different temperatures allow us to conclude that the temperature region where the maximum values of $|\Delta T_{AD}|$ are observed is near the LCST of the polymer PNIPAM, which allows us to conclude that the temperature of 36 $^0$C, which is in the range of physiological human temperatures, can be used as a working point for the experiment to demonstrate the activation of the polymer by a magnetic field. The magnetic field dependences of $|\Delta T_{AD}|$ demonstrate that the MCE reaches saturation above 4 T. This also indicates that the observed value of $|\Delta T_{AD}|$ is sufficient to induce a phase transition with the LCST of the PNIPAM polymer.

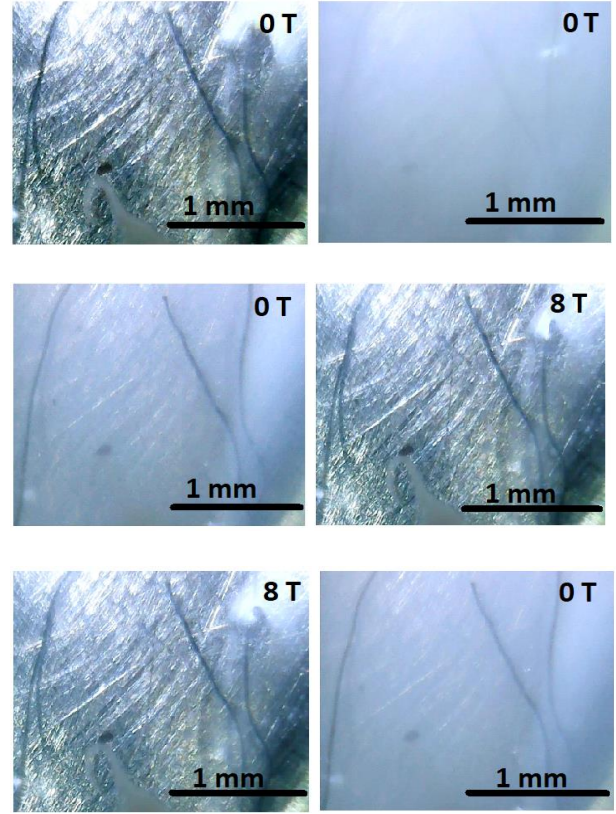

**Figure 5.** Optical images of activation LCST in PNIPAM by magnetic field in PNIPAM/FeRh composite: reference in zero field (a-b), «switch on» (c-d) and «switch off» modes (e-f); swollen hydrated state (a,d,e), swollen hydrated state, shrunken dehydrated state (e-f).

To demonstrate this effect, an in situ observation of the control of the polymer properties by a magnetic field due to MCE was carried out using an optical camera installed in a thermostabilized experimental cell, similar to the protocol used for PNIPAM/Gd [12]. Unlike Gd [29], the $Fe_{49}Rh_{51}$ alloy has inverse MCE, i.e., the polymer cools down when the magnetic field is switched on and vice versa. The *in situ* experiment to observe the control of polymer properties in the LCST region by magnetic field consisted of three consecutive steps and was organized according to the following protocol:

*Step 1*. Direct LCST phase transition with from swollen hydrated state to wrinkled dehydrated state without magnetic field by conventional heating;

*Step 2*. Reverse phase transition with LCST from wrinkled dehydrated state to swollen hydrated state moving in a magnetic field in as result of negative MCE («switch on»);

*Step 3.* Direct LCST phase transition from swollen hydrated state to wrinkled dehydrated state by extraction from the magnetic field as a result of the positive MCE («switch off» mode).

As noted, PNIPAM at 32 °C exhibits a transition from a swollen hydrated state to a wrinkled dehydrated state. In our experiments, this phenomenon can be visually observed with an optical camera as a transition of the polymer from a transparent to an opaque state (See video in Supplementary Information Section). For this purpose, the PNIPAM/FeRh composite sample was heated from 22 °C to 40 °C at a rate of 1.5 K/min in a zero magnetic field, and the change in the polymer's optical transparency in this range was clearly seen (*step 1*). As a reference, zero magnetic field images were obtained at hydrated 30 °C (Fig. 5 a) and dehydrated (Fig. 5 b) states of 36 °C for the PNIPAM/FeRh sample. As can be seen, moving the composite in 8 T magnetic field leads to its cooling and the reverse phase transition with LCST is induced, where the sample at 36 °C is in the wrinkled dehydrated state transitions to the swollen hydrated state as a result of negative MCE (*step 2*), which is clearly visible from the change in the transparency of the polymer (Fig.5 c-d). The positive MCE exhibits for FeRh, when the magnetic field is switched off, there is direct phase transition from with LCST (*step 3*), and the sample below the LCST of 32 °C changes from the swollen hydrated state to the wrinkled dehydrated state. If we compare the optical images taken for PNIPAN/Gd and PNIPAM/FeRh, we can observe that for PNIPAM/FeRh, the transition is more pronounced due to the fact that the FeRh magnetic phase transition temperature is closer to the PNIPAM LCST than that of Gd. Considering the experiment of controlling the properties of the thermoresponsive polymer by magnetic field in the PNIPAM/FeRh composite as successful, the next step of the experiment was to demonstrate the possibility of drug release via activation of the PNIPAM polymer by magnetic field as a result of negative MCE («switch on»). For this purpose, a PNIPAM/DOX/FeRh composite sample was used. The experiment for in situ observation of magnetic field-controlled doxorubicin release consisted of the following steps:

1. Setting the temperature to 40 °C in a thermally insulated sample chamber with the magnetic field switched off;

2. Injection of a fixed amount of water, slow thermostabilization, temperature setting of 36 °C; PNIPAM polymer is in a dehydrated state (optically opaque);

3. Moving the sample in magnetic field; induction of LCST transition of PNIPAM polymer to hydrated state (optically transparent); observation of drug release. To set up this experiment, a new experimental insert was designed and fabricated, which was an improved version of the insert used in the PNIPAM/Gd and PNIPAM/FeRh experiments [12,30].

The new version of experimental insert took into account the previous design flaws (non-rigid fixation of the sample, insufficient thermal stabilization, and thermal losses due to difficult control of the ambient volume). For this purpose, we first developed a CAD model of the lower part of the experimental insert with the specified parameters for the experiment's realization (sample dimensions, chamber diameter, diameter of the working gap of the magnetic system, etc.) (Fig. 6a). The frame of the model was printed on a 3D printer using the FDM method from PLA polylactide filament. The frame for fixing the object of study was printed under the dimensions of the PNIPAM/DOX/FeRh sample under study. A heater made of constantan wire Ø 0.1 mm was bifilamentally wound on the tube, which, together with a copper-constantan thermocouple glued to the sample holder body, performed a thermoregulation function to maintain a fixed temperature in the sample mounting region with an accuracy of 0.2 °C. After fixing the sample (Fig. 6b), the outer plastic casing was coaxially put on the inner plastic casing and fixed. The optical camera for observation of the experiment was led through the inner hole (1) of the outer plastic casing (Fig. 6 a).

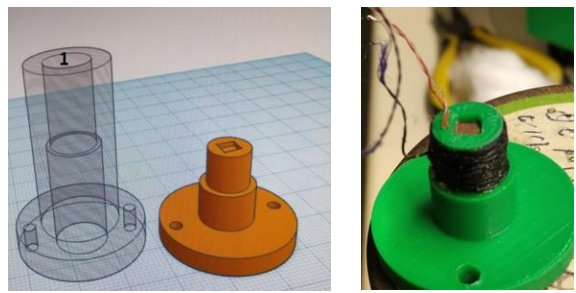

**Figure 6.** CAD model (a) and photo of the part of the experimental insert (b).

According to the step-by-step plan of experiment described above, the sample was first mounted and fixed, the camera was focused, and the initial temperature of 40 °C was set by the LakeShore 335 thermoregulator system. Next, the ~3.5 µL of distilled water with a temperature of about 40 °C was injected through a special hole using a medical syringe, followed by slow thermostabilization to 36 °C. This procedure was performed to avoid the premature release of doxorubicin by cooling the polymer to the LCST temperature. The starting temperature of 36 °C was chosen because of its proximity to human physiological temperatures. At this temperature, the PNIPAM polymer is in a dehydrated state but still retains the loaded doxorubicin. Next, the insert was inserted into the working region of the cryogenic magnetic system using a linear actuator, whereby the PNIPAM polymer was cooled to an LCST ~32°C, at which point doxorubicin release occurred. The magnetic field in the working area of the magnet was 3 T, which is comparable to the magnetic characteristics of modern commercial MRI systems (2–3 T). Figure 7. shows optical photographs captured in different modes before and after switching on the 3 T magnetic field, by which we can qualitatively conclude the realisability of the concept of DOX release by magnetic field activation of temperature responsive composite.

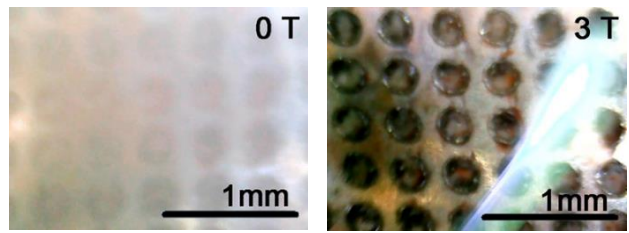

**Figure.7.** Captured photo before and after switching on a 3 T magnetic field from the surface of the PNIPAM/DOX/FeRh sample.

To verify the observed effect obtained by magnetic field activation, the aqueous solution of the DOX drug was collected from the surface of the sample with the help of a syringe and examined using a UV-visible spectrophotometer in transmission mode and a Raman spectrometer. To detect the drug dis-charge, the transmission spectra of the aqueous solution of doxorubicin collected from the surface of the sample after the experiment were taken.

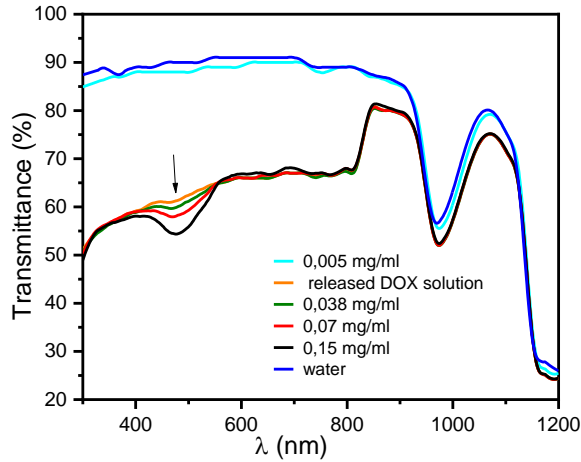

**Figure 8.** UV-VIS spectrophotometry spectra of DOX solutions.

Figure 8 shows the spectra recorded for doxorubicin in water solutions in transmission mode on the spectrophotometer. It is known that most drugs have intrinsic light absorption in the UV region, and some of them enter chemical reactions leading to the formation of colored compounds. This allows the development of new techniques based on the optical properties of compounds and aimed primarily at the determination of active components without their separation. Usually, at spectrophotometry, the absorption index or optical density is measured, which, according to Bera's law, is proportional under certain conditions to the concentration of the solution [31]. Their measurements allow to find the content of substances in solutions of unknown concentration. In a preliminary study of the absorption spectra of solutions, we found that doxorubicin has a pronounced minimum at 476 nm.

To quantify the dissolved drug, we prepared and measured the transmission spectra of solutions with a known concentration of doxorubicin. The results of spectrophotometric analysis show that the concentration in the sample under study is 0.014 mg/ml. Oscillatory spectroscopy, including Raman microspectroscopy, has been widely used over the past few years to explore potential biomedical applications [32]. Indeed, Raman microspectroscopy has been demonstrated to be a powerful, non-invasive diagnostic and monitoring tool. Figure 9 shows the Raman spectra of PNIPAM polymer, DOX, and released DOX solution in water. The characteristic CR peaks of DOX are mainly located at 440, 462, 989, 1082, 1209, 1241, 1299, and 1448 cm$^{-1}$. These characteristic peaks reveal extensive structural information about DOX molecules [33,34]. The 440 and 462 cm$^{-1}$ bands are attributed to the vibrations of CCO and CO bonds, respectively. The presence of these peaks related to DOX in spectrum 3 (*Released DOX*) indicates the stability of the released DOX.

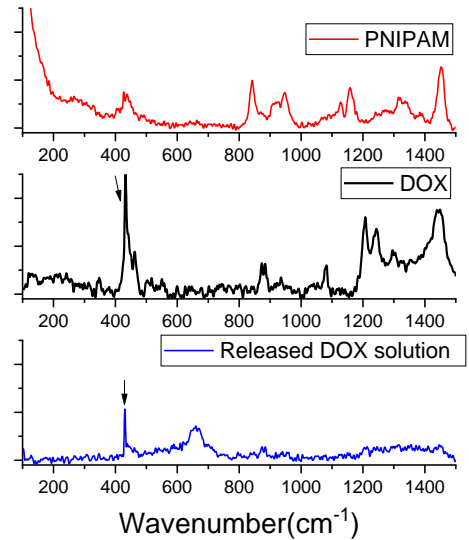

**Figure 9.** Raman spectra of PNIPAM, DOX, and released DOX.

To verify the biocompatibility of PNIPAM/FeRh composites, a series of tests were performed. In order to do this, PNIPAM/DOX/FeRh was cleaned of all doxorubicin and PNIPAM residues and recoated with a layer of PNIPAM polymer using the same method as before.

The MTT test showed that PNIPAM/FeRh had no cytotoxic effect, did not induce cell death, and reduced proliferative activity (Figure 10).

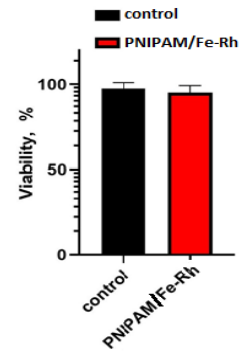

**Figure 10.** MTT assay of PMEF 72 h after cultivation on the surface of PNIPAM/FeRh scaffold.

Then, primary mouse embryonic fibroblasts were cultured on the same PNIPAM/FeRh sample used as scaffold, and their proliferation processes were investigated at 24, 48, and 72 hours. Morphology analysis shows that the scaffold provides PNIPAM/FeRh with good cell adhesion, cell spreading, and active proliferation during 72 hours of observation (Figure 11). Part of the adhered cells are localized directly in the wells, where they also have a spreading appearance, which confirms the high level of biocompatibility of the developed scaffold. It can be seen that some of the cells actively spread not only in the wells but also on the surface of the scaffold, and their cell morphology is characteristic of actively dividing fibroblast-like cells.

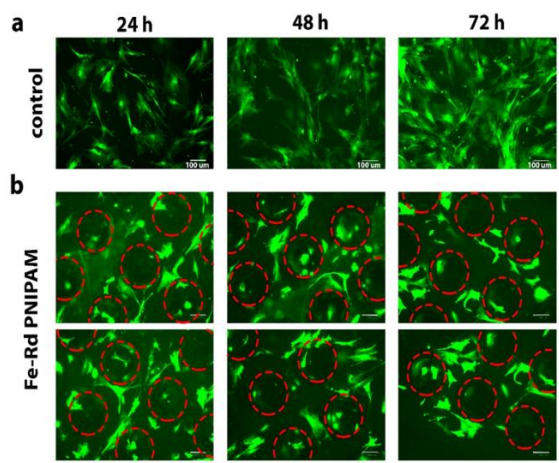

**Figure 11.** Cell morphology of primary GFP-mouse embryonic fibroblasts grown on the surface of a PNIPAM/FeRh-coated scaffold after 24, 48, and 72 h of cocultivation. Red lines indicate the cells on the scaffold for visualization of cells that are located directly in the wells. Control cells grown on cell culture plastic (polystyrene). Scale bar is 100 μm.

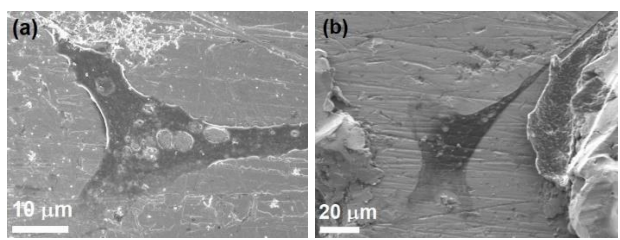

**Figure 12.** SEM images of primary mouse embryonic fibroblasts cultivated on PNIPAM/Rh-coated scaffold.

Figure 12 shows SEM images of individual cells cultured on the PNIPAM/FeRh surface. As noted, PNIPAM does not completely cover the laser-modified FeRh surface. As can be seen from the images, the cells are fixed and develop not only on the surface of the PNIPAM polymer (Figure 12 a) but also directly on FeRh (Figure 12 b).

## CONCLUSIONS

The concept of tuning of thermoresponsive polymer state by magnetic field in result of magnetocaloric effect were demonstrated using in situ experiments on PNIPAM/FeRh smart composite. The possibility to drug release loaded in PNIPAM/FeRh using as example doxorubicin were demonstrated in 3 T magnetic field, which can be reached commercial MRI sources. The DOX releasing were detected using UV-VIS spectrophotometry and Raman microspectroscopy. The PNIPAM/FeRh used as scaffold exhibits good biocompatible for mammalian cells and provides their effective adhesion and proliferative activity. The obtained results can be used for development of new magnetic field activated smart implants and theranostic systems.

### Supporting Information

The Supporting Information is available free of charge on the ACS Publications website.

Movie «Tuning of PNIPAM properties trough magnetocaloric effect induced in results of applied magnetic field changes in PNIPAM/FeRh smart composite» (AVI).

## AUTHOR INFORMATION

### Corresponding Author

* Abdulkarim Amirov (ORCID: 0000-0001-5311-2063), E-mail: amiroff_a@mail.ru

### Author Contributions

The manuscript was written through contributions of all authors. All authors have given approval to the final version of the manuscript.

### Notes

The authors declare no competing financial interest.

## ACKNOWLEDGMENT

This work was supported by the Russian Science Foundation (№ 24-19-00782). Authors are grateful to professor A.M. Tishin (M.V. Lomonosov Moscow State University, 119991, Moscow, Russia) for fruitful discussions and recommendations.